\documentclass{acm_proc_article-sp}[10pt]

\newif\ifpdf
\ifx\pdfoutput\undefined
   \pdffalse     
\else
   \pdfoutput=1  
   \pdftrue
\fi

\usepackage{times}

\ifpdf
\usepackage[pdftex]{graphicx}
\else
\usepackage{graphicx}
\fi

\ifpdf
\usepackage[pdftex,bookmarks=false,
            plainpages=false,naturalnames=true,
            colorlinks=true,pdfstartview=FitV,
            linkcolor=blue,citecolor=blue,urlcolor=blue]{hyperref}
\else
\usepackage[dvips]{hyperref}
\fi

\newcommand{\myhref}[2]{\ifpdf\href{#1}{#2}\else\htmladdnormallinkfoot{#2}{#1}\fi}

\conferenceinfo{OOPSLA '02}{2002 Seattle, WA, USA}
\CopyrightYear{2002}

\title{Monitoring and Debugging\\
  Concurrent and Distributed\\
  Object-Oriented Systems}
\numberofauthors{1}
\author{
\alignauthor Joseph R.~Kiniry\\
       \affaddr{Department of Computer Science}\\
       \affaddr{California Institute of Technology}\\
       \affaddr{Mailstop 256-80}\\
       \affaddr{Pasadena, CA 91125}\\
       \email{kiniry@acm.org}
}

\date{\today}

\begin{document}

\maketitle

\begin{abstract}

  
  A major part of debugging, testing, and analyzing a complex software
  system is understanding what is happening within the system at run-time.
  Some developers advocate running within a debugger to better understand
  the system at this level.  Others embed logging statements, even in the
  form of hard-coded calls to print functions, throughout the code.  These
  techniques are all general, rough forms of what we call \emph{system
    monitoring}, and, while they have limited usefulness in simple,
  sequential systems, they are nearly useless in complex, concurrent ones.
  We propose a set of new mechanisms, collectively known as a
  \emph{monitoring system}, for understanding such complex systems, and we
  describe an example implementation of such a system, called IDebug, for
  the Java programming language.

\end{abstract}

\category{D.2}{Software}{Software Engineering}
\category{D.2.5}{Software}{Software Engineering}[Testing and Debugging]

\terms{monitoring, debugging, assertions, tracing, logging, statistics,
  idebug, categories, levels}


%
\section{Introduction}
\label{sec:Introduction}

Modern application development is extremely complicated.  Today's
developers, as members of large teams, are building desktop applications
with millions of lines of code.  Large-scale applications must be built by
judicious use of existing code and ideas: code and design reuse,
compositional architectures, patterns, and other similar models are the
modern tools of our trade.

\subsection{Modern Methods, Dark Ages}

But how do most developers do development and testing?  While the majority
of developers use advanced integrated modeling and development environments,
they also use a smattering of low-tech print statements for system
monitoring and debugging.

What is wrong with this scenario?  
\begin{enumerate}
\item It is an undisciplined development methodology: there is little or
no relationship between application requirements and monitoring, debugging,
and test code.  
\item There is no relationship between a component's specification and the
  component test code.
\item Test code is manually embedded in the code that is sometimes shipped,
  leaving a host of deployment issues to deal with at delivery-time.
\item Output is unstructured and often unparseable.  Finally, there is no
  easy way to redirect or log test output to any destination other than a
  pipe or a file.
\end{enumerate}

While ``archaic'' languages like C and Fortran have an assertion mechanism,
surprisingly, some modern popular languages are designed without regard for
system debugging and some do not support even simplistic debugging
constructs like assertions.

\subsection{A Specific Example: Debugging in Java}

For example, the Java programming language provides very few built-in
constructs for debugging classes, components, and systems.

Typically, a Java programmer relies upon language features and development
tools for debugging.  Java provides array bounds checking, static type
checking, variable initialization testing, and exceptions to assist in code
debugging.  While programming environments provide sophisticated
source-code debuggers, most developers seem fixated on using primitive
\texttt{println}'s to debug their code.

Java is missing several traditional core debugging constructs as well, the
most critical of which is \emph{assertions}\footnote{With the release of
  JDK 1.4 we now have an \texttt{assert} keyword, nearly seven years after
  the initial version of Java.}.  Typically, if an assertion is violated, a
program is aborted.  In modern object-oriented systems we often need to
have options other than halting the program execution (e.g, throwing an
exception).

\subsection{Filling the Holes}

IDebug has been designed to ``fill the holes in Java''.  It is a monitoring
framework composed of a set of components that provide fundamental
monitoring and debugging constructs like assertions, error messages,
logging, and more.  Applications and/or components using IDebug have a
unified, manageable, flexible and extensible interface for monitoring and
debugging.

The point of this paper is not to describe IDebug in full detail.
Instead, our primary focus is to discuss what we believe are the
fundamental aspects of such a monitoring and debugging subsystem,
particular for concurrent or distributed systems.  We hope to convince the
reader that the complex applications and components being developed today
\emph{necessitate} a new view on monitoring, debugging, and testing
frameworks.  In particular, with the range of application domains
assaulting the average developer (everything from embedded to distributed
systems), a comprehensive and flexible monitoring, debugging, and testing
subsystem is \emph{essential}.

\section{Background}
\label{sec:Background}

Programming technologies have evolved rapidly over the years. New
programming models have emerged, new languages have gained popularity, new
tools have been adopted, and yet, several core monitoring and debugging
constructs have changed very little.

The primary constructs for monitoring and debugging are the \emph{call
  stack}, the \emph{logging message}, and the \emph{assertion}.  A
\emph{call stack} is the sequence of nested function or method calls of a
specific execution context.  The \emph{logging message} is a structured
message that logs special events (errors, warnings, etc.).  They are
usually time-stamped, have priority levels, and are organized according to
some taxonomy.  Finally, an \emph{assertion} is a boolean predicate that,
if false, indicates an error in program code.


Since IDebug is an object-oriented monitoring and debugging framework, a
review of object-oriented debugging and frameworks is in order.

\subsection{Object-Oriented Debugging}

Debugging object-oriented programs is not the same as debugging
procedural or functional programs.  Because most object models enforce
modularity and encapsulation, one must test the implementation
\emph{and} the interface of a class or module.

A specification of an interface is called a
\emph{contract}~\cite{HamiltonPowellMitchell93, HelmHollandGangopadhyay90,
  Holland92, Meyer88}.  A class's contract specifies the externally visible
behavior that a class guarantees.  Contracts are typically specified via
three constructs: \emph{pre-conditions}, \emph{post-conditions}, and
\emph{invariants}.  Using these three constructs, many of the safety
properties of a class can be specified\footnote{Progress conditions can
  also be specified in contracts but we only focus on safety conditions
  here.}.

\subsection{Frameworks}

A \emph{framework} is a collection of programming constructs (e.g., classes,
components, interfaces, etc.) that provides a unified model and interface
to a specific set of functionality.  Frameworks are integrated into a
system via inheritance and client relationships.

A framework in Java is typically implemented as a collection of classes
organized into one or more \emph{packages}.  IDebug is just such a
framework since it is implemented as a set of JavaBeans and other classes
collected into the package \texttt{idebug}.  These classes are used either
(\emph{a}) as ``normal'' classes with standard manual debugging techniques,
or (\emph{b}) as components within visual JavaBean programming tools.

\section{Constructs}
\label{sec:Constructs}

The core set of monitoring and debugging constructs: the assertion, call
stack, and logging message, are provided by simple language constructs.

Assertions are simply provided via method calls.  When an assertion fails,
one of the following takes place: a specific exception is thrown, the
current thread is stopped, the current context is halted (e.g., an applet or
servlet is stopped), or the system is halted.

Call stack introspection is provided as part of the Java language
specification.  The call stack is used in several ways by the functionality
introduced in the next few sections.

Logging messages are simple formatted text strings that are sent to one of
several \emph{output channels}.  An output channel is simply any data
channel through which we can send logging messages.  Examples include the
Java console, a \texttt{String\-Buffer}, a file, etc.

This basic set of constructs is only the beginning of the story.  They are
generally only sufficient for simple, sequential, non-distributed systems.
We augment this set by several new constructs specifically for concurrent
and distributed systems based upon requirements imposed by development (how
software is written) and system (how software is executed) contexts.

\subsection{Functionality for Complex Systems}

The development of complex systems often involves dozens to hundreds of
developers located at several sites.  This situation necessitates a
decomposition of responsibilities in the development process.

Collaboration between teams and individuals working on different parts of
the system is often ad hoc and opportunistic.  Thus, a system-wide,
regimented, fixed monitoring, debugging, and testing process is out of the
question.

As a result, development groups define their own terminology and priority
structure specific to their problem domain.  The definition of such
structure is accomplished with \emph{levels} and \emph{categories} in
\emph{contexts}.

\subsubsection{Levels}
\label{sec:levels}

Priority structures are realized by \emph{levels} that let a developer
assign a priority to actions.  Levels organize debug information into a
totally ordered set.  The default levels range from 1 to 9 with
well-defined increments: NOTICE (1), WARNING (3), ERROR (5), CRITICAL (7),
and FAILURE (9).  Localized error messages are associated with these
standardized levels for regular logging output.  The range of levels, like
all other customization, is refined on a per-context basis (see below).
For example, for a more complex system with subtle failure modes, perhaps a
range of 1 to 100 is more appropriate.

\subsubsection{Categories}

Terminology is realized by an ontology of \emph{categories}, strings like
\texttt{"NETWORK"} and \texttt{"GARBAGE\-\_COLLECTOR"} that denote the
subsystem correlated with a specific debugging action.  Each category has a
level associated with it as well for filtering purposes.

At runtime, both categories and levels are used to prune information
according to the demands of the current execution.  This mechanism is
discussed in detail in the next section since it interplays with the
concurrency constructs.

\subsubsection{Context}

Additionally, subgroups of large development teams often have independent
process and practices.  But, especially during integration testing and
system maintenance, they often need to understand, share, and utilize a
codification of these practices.  To satisfy this need we introduce the
first-class notion of a \emph{context}.

A \emph{context} is an object that captures the full monitoring, debugging,
and testing context of a subsystem.  Independent ontologies (set of
categories), level sets, filtering (see the next section), and more are
bound to a context.  This context is saved and restored as needed and is
exchanged among teams like any other development artifact.

\subsubsection{Statistics}

Finally, a major challenge after constructing a complex system is
understanding what it is actually doing as it runs.  This understanding
goes a long way toward helping with system evolution, maintenance, and
optimization.

We introduce a \emph{statistic} construct to assist with exactly this
problem.  Each statistic has a unique identifier and a description.  It
also has a unit (e.g., meters, frames per second, etc.), a scaling factor,
initial values, and default increment and decrement values.  We can also
arbitrarily manipulate (reset, set, increment and decrement) a statistic.
Finally, each statistic has report generation facilities specific to its
domain.

For example, suppose we needed to gather statistics on a message passing
system.  We define a statistic called \texttt{Msg\-Per\-Second} whose units
are \texttt{"messages per second"}, scale is \texttt{1000}, default value
is 0, and default increment and decrement are both 0.001.  This means that
this statistic is initialized to zero and each time the statistic is
incremented it gains 0.001 * 1000 = 1 message per second.  The information
on units is used when generating statistics reports.

\subsection{Functionality for Concurrent Systems}

Monitoring and debugging concurrent systems necessitates the introduction
of new concepts specific to concurrent environments.

Concurrent systems have multiple threads of control.  Typically, each
thread of control accomplishes some specific task: it transfers data,
refreshes a GUI, etc. Thus, each thread is potentially the jurisdiction of
a different development team.

Also, concurrent systems often have a hierarchical arrangement for threads
of control.  Java's \emph{thread groups} are one example.  Threads are
collected into named groups for security, management, and control purposes.

Therefore, we ``tune'' functionality on both a per-execution thread and a
per-thread group basis.  Each thread and each thread group has a context.
These contexts are exactly the same as the contexts discussed earlier
except they are bound to a single thread of execution or a thread group of
such threads.

Contexts can be manipulated in a concurrency-safe fashion at runtime to
dynamically change the monitoring behavior of an application.  The primary
manipulations of such contexts are changes in their filter specifications.

\subsubsection{Tuning}

Contexts can be tuned in several ways.  Monitoring is tuneable on a:
\begin{enumerate}
\item global basis by turning monitoring on or off as a whole, independent
  of any other settings.
\item per-class basis.  Individual classes can be identified as being
  important or unimportant to monitoring.
\item threshold basis using monitoring levels.  A \emph{current monitoring
    level} can be set, after which time all monitoring code that is
  annotated with at least this level is evaluated.
\item per-category basis.  Specific categories can be identified as being
  important or unimportant to monitoring.
\end{enumerate}

For all of the above functionality, the monitoring system uses the call
stack to support runtime-configurable filters for logging messages based
upon the current execution context of a thread.

\begin{figure}[h]
  \begin{center}
    \ifpdf
    \includegraphics[width=3in]{tuning_idebug}
    \else
    \includegraphics[width=3in]{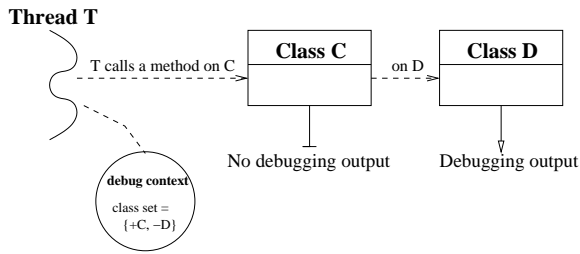}
    \fi
    \caption{Tuning IDebug at run-time}
    \label{fig:tuning_idebug}
  \end{center}
\end{figure}

Consider Figure~\ref{fig:tuning_idebug}.  If a thread $T$ specifies that it
is interested in a class $C$ but not a second class $D$, then monitoring
statements in $C$ will be executed when $T$ is inside of $C$, but
monitoring statements in $D$ will be ignored.  These contexts are saved to
persistent storage, thus ``named'' special-purpose contexts are created for
reuse across a development team to help support and enforce a monitoring
process.

Additionally, these \emph{concurrent contexts} can be shared across threads
and thread groups.  A shared context that is tuned immediately impacts all
execution constructs to which it is bound.

\subsection{Functionality for Distributed Systems}

The framework supports several extensions for monitoring distributed
systems.

\begin{figure}[h]
  \begin{center}
    \ifpdf
    \includegraphics[width=2in]{call_currying}
    \else
    \includegraphics[width=3in]{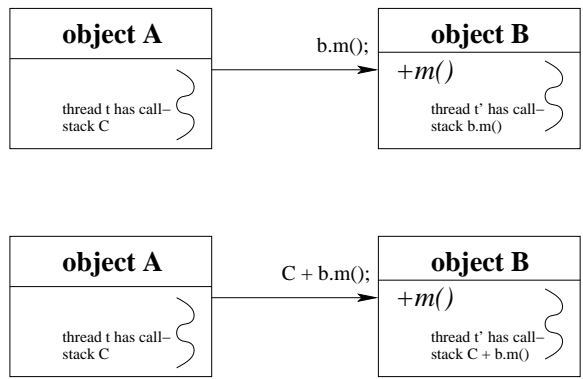}
    \fi
    \caption{Call stack currying}
    \label{fig:call_stack_currying}
  \end{center}
\end{figure}

One problem typical of monitoring distributed systems is a loss of
execution context when communication between two non-local entities takes
place.  

Consider the top-half of Figure~\ref{fig:call_stack_currying}.  When object
$A$ invokes a method $m$ on object $B$, the thread within $m$ does not have
access to the call stack from the calling thread in $A$.  The IDebug
package supports what we call \emph{call stack currying} to solve this
problem.  Information such as source object identity, calling thread call
stack, and more is available to the monitoring framework on both sides of a
communication.  Such information is curried across arbitrary communication
media (sockets, RMI, etc.), as seen in the bottom-half of the figure.

\begin{figure}[h]
  \begin{center}
    \ifpdf
    \includegraphics[width=3in]{mobile_agent}
    \else
    \includegraphics[width=3in]{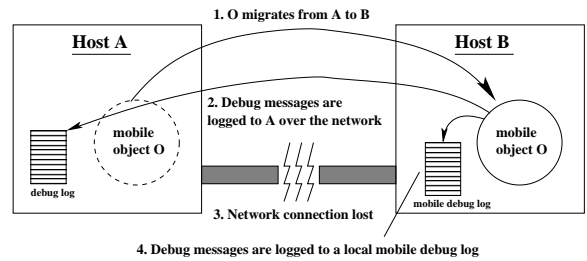}
    \fi
    \caption{Mobile agent system monitoring}
    \label{fig:mobile_monitoring}
  \end{center}
\end{figure}

The IDebug package also supports the monitoring of mobile agent systems.

Mobile agent architectures often support disconnected computing.  For
example, in Figure~\ref{fig:mobile_monitoring}, (1) an object $O$ migrates
from machine $A$ to machine $B$.  For some time (2), $O$ can log debug
information back to host $A$, but then (3) $B$ might become disconnected
from the network.  Since $B$ can no longer communicate with $A$, and
printing monitoring information on $B$'s display is not useful, or perhaps
even possible, $B$ must log monitoring information for later inspection
(4).  

To support this functionality, the IDebug package provides serializable
debug logs.  These logs are carried around by a mobile object and inspected
later, perhaps when the object migrates back to its original host.

\section{Impact on System Design and\\Development}
\label{sec:Impact_on_System_Design_and_Development}

The impact of a monitoring framework like IDebug to systems design and
development is extensive.
\begin{itemize}
\item \emph{Analysis.} The assumption of the availability of a flexible and
  comprehensive monitoring framework results in a conscientious domain
  analysis from a monitoring, debugging, and testing perspective.
  
  Normally, this type of analysis either (\emph{a}) never happens or
  (\emph{b}) is rolled in late to system design or development.  An example
  of such analysis is highlighted in
  Section~\ref{sec:Application_Requirements}.
\item \emph{Design.} Design schedules are shorter than otherwise because
  the monitoring subsystem does not have to be designed by scratch as an
  afterthought, as is typical in many design efforts.  Likewise, a new
  focus is found with respect to design integration of monitoring,
  debugging, testing, deployment, and management.  Standardized
  methodologies and semantics for monitoring are considered early in system
  design.
\item \emph{Process.} With the availability of such a monitoring
  framework, especially when coupled with complementary tools like
  JPP~\cite{KiniryCheong98} and the CDL parser~\cite{KiniryCDL99},
  the development process becomes much more focused on specification
  and rigorous testing.  Likewise, component and system documentation
  is more organized, uniform, and complete.
\item \emph{Testing.} In our experience, components and systems built with
  such a framework are \emph{much} more robust.  More thought has been put
  into monitoring, debugging, and testing because of the process imposed by
  the use of such tools.
\item \emph{Deployment.} Since monitoring functionality is tuneable at
  runtime, some debug code can be shipped with product at no loss in
  performance, though at the cost of slightly larger executable size.
  System monitoring, debugging, and testing in the field is
  straightforward, empowering end-users' contributions to the process, and
  potentially lowering support costs.
\end{itemize}

\section{Requirements}
\label{sec:Requirements}

In this section we will briefly present our project analysis, including our
project concept dictionary, a review of our requirements for the monitoring
package, and our goals.

\subsection{Project Dictionary}

At the beginning of the project analysis phase, a dictionary of concepts
was developed so that all designers, developers, and users would have a
clear and common language.  The dictionary of terms is available in the
full, high-level EBON specification of the system included with the package
and available
\myhref{http://www.kindsoftware.com/products/opensource/IDebug/docs/BON/}
{online}.

\subsection{Core Requirements}
\label{sec:core}

We require that the IDebug framework support the following
requirements.  The framework must:
\begin{enumerate}
\item \emph{Provide an assertion mechanism.}  Assertions are the core
  construct of any debugging system.  Assertions are inserted in program
  code either manually or automatically, and, if an assertion is violated,
  an error message is logged and/or a runtime exception is thrown and the
  program (potentially) halts.
\item \emph{Support the output of logging messages.}  Printing
  miscellaneous monitoring messages, perhaps outside the context of the
  primary interface of a component, is essential in a good monitoring
  suite.
\item \emph{Support multiple monitoring levels.} Different types of
  errors, messages, and situations require different levels of
  response.  An adequate monitoring framework should not only support
  levels, but the set should be ordered so that user- or
  developer-tuneable filtering of debug output can take
  place\footnote{An external filtering mechanism like \texttt{grep}
    could be used instead, though is usually more tedious for the
    tester.}.
\item \emph{Complement the standard Java exception mechanism.} Since this
  is a monitoring framework built for the Java language, it should work
  with, not against, the built-in exception mechanisms.  In particular,
  prudent use of exception types (\texttt{Exception} verses \texttt{Error})
  is necessary so that the framework is not overly intrusive to the
  developer\footnote{For example, if all exceptions were runtime
    exceptions, the developer would have to bracket nearly all code with
    \texttt{try-catch} blocks.}.
\item \emph{Work with all development environments.}  IDebug must work
  with everything from the most flashy IDE to the lowly CLI runtime.
  This means that IDebug must be implemented as ``100\% Pure Java'';
  no proprietary extensions or native code may be used.
\end{enumerate}

\subsection{Application Requirements}
\label{sec:Application_Requirements}

Because our group builds a wide range of Java applications and components,
we needed IDebug to support monitoring with all types of Java programs.
This means that the framework must provide debug functionality that
complements the following application types.  Each type of application
listed below is followed by a (non-unique) implication of that particular
application assumption.  In other words, each different domain implies the
necessity of new piece of functionality.
\begin{enumerate}
\item \emph{Console-based applications.} Sometimes we want to send logging
  messages to an output stream different from our system's \textsc{stdout}
  or \textsc{stderr} streams.
\item \emph{Graphical user interface applications.} Occasionally, we want
  to send logging messages to independent windows or message sub-frames
  within a larger application.
\item \emph{Console-less applications.} If there is no output channel,
  logging messages for later analysis is a reasonable course of
  action.
\item \emph{Independent components such as beans, servlets, doclets, etc.}
  Independent components should be able to maintain independent semantics
  and contexts.  Conversely, sometimes it is useful to have a compositional
  application share a debug context among its components.
\item \emph{Mobile agent/object applications.} If an application has
  mobile sub-components, their debug contexts need to be mobile as
  well.  Additionally, logging message output and/or storage should
  be location-independent and location-aware.
\item \emph{Distributed applications.} Distributed applications imply (at
  least) distributed control, distributed context, and distributed logging.
\end{enumerate}

\subsection{Innovative Requirements}
\label{sec:Innovative_Requirements}

Finally, we wish to support a set of (what we consider) ``innovative''
capabilities.  While most of these goals are independent of the target
language, they are facilitated by many of Java's more advanced features.
The list of innovative requirements includes support for:
\begin{enumerate}
\item \emph{Categorized monitoring.} Monitoring messages, errors, warnings,
  etc. should not only have a value (the debug level), but should have a
  category (a classification or taxonomy).
\item \emph{Per-class tuning.} A developer should be able to selectively
  turn monitoring on or off at a per-class level.
\item \emph{A configurable runtime.}  We should not force developers to
  adopt our monitoring semantics.  New semantics (ranges, base categories,
  etc.) should be configurable at design- and run-time.
\item \emph{Multiple output interfaces.} All logging messages need not be
  sent to the same output channel.  Consider messages generated by UNIX's
  \emph{syslog} facility.  Some messages are sent to the console, some
  are logged in a file, and some are sent directly to the system
  administrator via email.
\item \emph{Concurrent monitoring.} Each thread and thread group within a
  runtime should be able to construct its own context.  More precisely,
  most of the above configurable options (categories, classes, semantics,
  output interface, and level) should be configurable on a per-thread and
  per-thread group basis.  Additionally, these options should be
  configurable at runtime.
\item \emph{Persistent contexts.} Once a context is created, it should be
  possible to save it to storage for later access.  This way, contexts can
  not only be shared across sets of components, but they can be shared
  across groups of developers.
\item \emph{Statistics gathering.} Gathering information about key system
  aspects over a long time-frame helps us understand critical bottlenecks
  and hot-spots in our systems.  This information can also help the system
  \emph{self-tune}, changing resource utilization or operating practices at
  run-time depending upon the current situation.
\end{enumerate}

Now that we have a common vocabulary and understand the problem domain
and the design goals, we will consider IDebug's design.

\section{Design}
\label{sec:Design}

Due to space restrictions, we will only briefly cover the design of IDebug
in this section.  We will describe IDebug's organization, its primary
subsystems, and give an example of its use.  Readers interested in more
information should see the IDebug home page, available via
\myhref{http://www.kindsoftware.com/products/opensource/IDebug/}
{KindSoftware}.

Several group members helped with IDebug's initial analysis and design in a
whiteboard-brainstorming session.  Design refactoring first was done
``by-hand'', then was moved to the Java design tool
\myhref{http://www.togetherj.com/} {Together/J}.

\subsection{Framework Structure}

IDebug's primary interfaces are called \texttt{Debug\-Output} and
\texttt{Debug\-Constants}.
\begin{description}
\item[Debug\-Output.] This is the interface to output methods that are used
  to send logging messages to various output channels.  Features include
  methods like \texttt{printMsg}, \texttt{print}, and \texttt{println}.
\item[Debug\-Constants.] This is the interface that collects the semantics
  of the package including monitoring level ranges, standard logging
  messages, etc.  It can be extended to change these values for specific
  monitoring sub-packages, applications, etc.  An example of such a subtype
  is included as \texttt{idebug.\-examples.\-French\-Constants}. A set of
  default categories are specified in the specification of this interface,
  as discussed in Section~\ref{sec:levels}.
\end{description}

The primary classes of note are \texttt{Debug}, \texttt{Assert}, and
\texttt{Context}.
\begin{description}
\item[Debug.] Debug is the core class of IDebug's monitoring facilities.
  The Debug class is used to configure monitoring for a component.  Debug
  has methods for configuring the following options:
  \begin{itemize}
  \item global, per-thread, and per-thread group monitoring activation
    (i.e., whether \emph{any} debugging predicates are checked or
    monitoring commands are executed),
  \item global debug output interface, debug semantics, and debug
    levels,
  \item categories and class-level tuning at the global, per-thread, and
    per-thread group level,
  \item configure contexts (see \texttt{Context} below) on a per-thread and
    per-thread group basis.
  \end{itemize}
\item[Assert.] The class used to make assertions.  A reference to an
  \texttt{Assert} object is obtained by calling the
  \texttt{get\-Assert} method of the \texttt{Debug} class.  Assertions
  are made in program code by calling the \texttt{Assert.\-assert}
  method which has several polymorphic forms.  The class also provides a
  static interface to making assertions.
\item[Context.] This class is the data structure that contains the
  information relevant to monitoring on a per-thread and a per-thread group
  basis.  Output interface, semantics, categories, class-level tuning, and
  current level are all configured with this class.
\end{description}

Example extension classes and blackbox test code are provided with IDebug;
they are collected in the \texttt{idebug.\-examples} package.  In
particular, the \texttt{French\-Constants} interface implements the
\texttt{Debug\-Constants} interface and is an example set of alternative
debug semantics where error messages are in French and debug ranges are
integers ranging from 1 to 100.  The \texttt{Debug\-Tests} class is the
main class for all test code; \texttt{Test\-Suite\-Thread} contains the
actual blackbox test code.

\subsection{Framework Behavioral Description}

The full behavioral specification of IDebug is included with the package as
(a) a set of UML state, collaboration, and sequence diagrams, and (b) a
full Extended BON (EBON) specification~\cite{EBON01}.  We will not provide
these diagrams here due to space restrictions.  Instead, we will summarize
how one interacts with the IDebug framework from a developer's point of
view.

Readers should refer to Appendix~\ref{app:Example_Code} for example code
using IDebug.  Line numbers referenced below refer to the lines labeled in
this example code.  These code examples are from
\myhref{http://www.jiki.org/} {Jiki}, an open source collaborative web
architecture based on distributed components, thus the feature naming
conventions.  Note that almost all comments have been removed from the
example code and it has been edited for relevant content.

\subsubsection{Usage Overview}

\texttt{idebug.\-Debug} is the core class of the IDebug monitoring
facilities.  The \texttt{Debug} class is used as the central facility for
configuring monitoring for a component or application (lines 3, 12, 13,
15).  All logging commands, on the other hand, are handled in the
\texttt{Debug\-Output} classes (lines 5, 14, 15, 25-26).  Finally, all
assertions are handled in the \texttt{Assertion} class (lines 4, 13,
23-24).

\paragraph{Core Configuration and Global Options.} 
The \texttt{Debug} class is non-static\footnote{See
  Section~\ref{sec:Design_Decisions} below for a discussion of this design
  choice.}.  The first step a component or application must take is the
construction a new instance of a \texttt{Debug} object (line 12).  If an
alternate implementation of monitoring semantics (i.e., categories, levels,
error messages, etc.) is needed, the implementation is passed a
\texttt{Debug\-Constants} interface via a constructor of \texttt{Debug}
(lines 11-12).

\paragraph{Per-Thread and Per-Thread Group Contexts.}
Each thread needs to construct a debugging context to detail its specific
debugging needs.  After creating a valid debugging context, encapsulated in
the \texttt{Context} object, this object is passed to the instance of
\texttt{Debug} via the \texttt{add\-Context} method so that the debugging
runtime system has a record of the thread's context.

Note that the debug runtime keeps a reference to the passed
\texttt{Context} object, it does not make a copy of it.  Thus, you can
modify the \texttt{Context} (change debugging levels, add new
thread-specific categories, etc.)  after the context is installed and
changes will be noted immediately by the debug runtime\footnote{A
  discussion of this design choice is detailed below. Note also that our
  code sample uses the global debug context, configured implicitly in lines
  12, 15, and 16.}.

\paragraph{Debug Output Configuration.}
Finally, the output medium of the debugging runtime has to configured.
This is accomplished by constructing an implementation of the
\texttt{Debug\-Output} interface, e.g., \texttt{Console\-Output}.  This
object is then passed to the \texttt{Debug} object via the
\texttt{Debug.set\-Output\-Interface} method (lines 14-15).

\paragraph{Usage.}
The IDebug framework is now fully configured.  A call to
\texttt{debug.get\-Assert} at any time returns a reference to the debug
run-time's \texttt{Assert} object (line 13).  If a non-default
implementation of \texttt{Debug\-Constants} interface is not installed, a
call to \texttt{debug.\-get\-Debug\-Constants} returns a reference to these
debug constants.  Our code sample installed its own version of this
interface, thus this call is unnecessary.

Finally, one calls the various methods of \texttt{Assert} and
\texttt{Debug\-Output}.  The \texttt{assert} method of the \texttt{Assert}
object is used to make code assertions (lines 23-24; note the mapping
between the specification of the \texttt{doGet} method in lines 19-20 and
the corresponding assertion).  The \texttt{print}, \texttt{println}, and
\texttt{printMsg} methods of the \texttt{Debug\-Output} instance are used
to output logging messages (lines 25-26).  Additionally, methods like
\texttt{Utilities.\-dump\-Stack} can be used to perform full stack dumps.

\subsection{Design Decisions}
\label{sec:Design_Decisions}

Several non-trivial decisions were made during the design and
implementation of IDebug.  Some of these decisions are summarized below.

\subsubsection{Static Verses Dynamic Interface}

The original implementation of IDebug had a \texttt{Debug} class that was
completely static.  Meaning, all methods of \texttt{Debug} were declared
\texttt{static} so that it behaved more like the interface to a library
than an object in a framework.  Since an instance of \texttt{Debug} didn't
have to be constructed, a reference to the object did not need to be handed
around to various subsystems of the larger system being monitored.

We found that there were several drawbacks to this approach.  First,
changes to the interface of \texttt{Debug} necessitated a potentially large
set of changes to program code.  Second, most systems have a shared
debugging context across subcomponents, so even though a reference to an
instance of \texttt{Debug} didn't have to be shared across components, an
instance of \texttt{Context} often did.  This restriction destroyed
the whole reason for making \texttt{Debug} static in the first place.

Since moving to a non-static design, we have come across a few restricted
situations where a static interface would be useful.  Thus, the next
release of IDebug will likely support both methodologies.

\subsubsection{Persistence}

The \texttt{Context} class is the only class in the IDebug package
that implements the \texttt{Serializ\-able} interface.  We decided that
centralizing IDebug's configurability in a single class would facilitate
configuration reuse.  These contexts can be saved to storage and reused
across application executions, project teams, or different component design
and development efforts to standardize the design and execution of the
testing process.

\subsubsection{Runtime Configurability}

As mentioned previously, changes to the state of an installed
\texttt{Context} object has instant effect on the debug runtime.
Thus, we provide a \texttt{clone} method so that multiple identical base
contexts can be used across a set of threads or thread-groups for
independent configurability.  Note that if a single context is installed
for many threads, changes to its state instantly affect all related
thread's debug contexts.

\subsubsection{Extensibility}

To simplify framework extension we designed two orthogonal interfaces for
extensibility: \texttt{Debug\-Output} and \texttt{Debug\-Constants}.

If a new output interface for logging messages and assertions is needed, a
developer simply implements \texttt{Debug\-Output}.  See
Section~\ref{sec:Future_Work} for ideas about such extensions.  New
semantics can be configured by implementing the \texttt{Debug\-Constants}
interface.  See the next section for details on how semantics are refined.

\subsubsection{Class-Specific Debugging}
\label{sec:class-spec-debugg}

A decision was made to make all class-specific debugging configuration
additive \emph{and} reductive.  One can either remove all classes from the
debugging table then add classes one by one, or one can add all potential
classes then remove them one by one.  Meaning, when one adds ``*'' (a
wildcard indicating that \emph{all classes} should be added to the debug
context), one is \emph{not} adding all classes \emph{currently defined in
  this VM}; one is adding \emph{all classes currently defined and all
  classes that might ever be defined} in this VM.  See
Section~\ref{sec:Future_Work} for a discussion of future work along these
lines.

\subsubsection{Context Configurability}

As mentioned previously, debugging options should be configurable on a
per-thread or per-thread group basis.  On further consideration, we decided
that two configurable settings should \emph{not} be switchable at runtime:
debug semantics and output interface.

The reason for this decision might not be immediately obvious, but consider
the following two points:
\begin{itemize}
\item Debugging output might be queued due to the temporary unavailability
  of an output channel or user, and
\item Source code that uses a debugging package makes explicit assumptions
  about the semantics of the package.  Meaning, while debugging semantics
  might be switchable at runtime by the \emph{framework}, it is not
  (usually) switchable at runtime for the application \emph{using} the
  framework.
\end{itemize}

Due to these factors, the configuration of debugging semantics and output
interface is immutable.  Meaning, once these options are set for a
debugging context, they cannot be changed.

Note that a new context can be \emph{created} and installed.  All the other
flexibility mentioned in Section~\ref{sec:Innovative_Requirements} is fully
configurable at runtime on a per-thread and per-thread group basis.

\subsection{Framework Extensibility}

The IDebug framework is extensible in two dimensions: \emph{debug
  semantics} and \emph{output interfaces}.

\subsubsection{Framework Semantics}

The semantics of the package can be changed by implementing new versions of
\texttt{Debug\-Constants}.  An example of such an extension is provided in
the form of the \texttt{French\-Constants} class in the
\texttt{idebug.\-examples} package.  This class provides an implementation
of \texttt{Debug\-Constants} that differs from the default implementation
(\texttt{Default\-Debug\-Constants}) in two ways:
\begin{enumerate}
\item Debug levels range from 1 to 100 instead of 1 to 10, 
\item Default debugging levels have been adjusted for this new
  granularity of debug levels, and
\item Default logging messages, categories, and documentation are
  provided in French.
\end{enumerate}

\subsubsection{Output Interfaces}

New implementations of \texttt{Debug\-Output} can be designed to support
sending logging messages to alternative output media/channels.  The
framework comes with several implementations: \texttt{Console\-Output},
which sends messages to the console of a Java runtime;
\texttt{Writer\-Output}, which sends messages to a \texttt{Writer} which
can be used as part of a normal \texttt{java.io} compositional data stream;
\texttt{Window\-Output}, which sends messages to a Swing window; and
\texttt{Servlet\-Log\-Output} which sends messages to a servlet logging
interface.

Now, we'll briefly discuss the implementation of the IDebug framework.

\section{Implementation}
\label{sec:Implementation}

As mentioned previously, IDebug is implemented as a collection of Java
classes organized into two packages.  IDebug is shipped as either a
Jar or Zip file with full documentation, formal specification (UML and
EBON), user's guide, and more.

\subsection{Size and Performance}

\subsubsection{Implementation Size}

\begin{table}[htbp]
  \begin{center}
    \begin{tabular}{|l|l|}
      \hline
      \multicolumn{2}{|c|}{Implementation Summary} \\
      \multicolumn{2}{|c|}{(with test and example code)} \\ 
      \hline\hline
      \textbf{Total Number of Packages} & 3 \\ \hline
      \textbf{Total Number of Classes} & 20 \\ \hline
      \textbf{Total Number of KB of Java} & 187.7KB \\
      (includes code, documentation, & \\
      and whitespace) & \\ \hline
      \textbf{Total Number of KB of classfiles} & \\
      Independent class files & 58KB \\
      Jar (compressed) format & 32.1KB \\ \hline
      \textbf{Total Number of Lines of Code} & 3808 \\ \hline
      \textbf{Total Number of Lines of Comments} & 2639 \\ \hline
      \textbf{Comments/Code} & 69\% \\ \hline
    \end{tabular}
    \caption{Implementation Summary}
    \label{tab:impl_summary}
  \end{center}
\end{table}

The implementation size of IDebug is summarized in
Table~\ref{tab:impl_summary}.  It is clear that, since the whole of IDebug
is around 32KB of bytecode, it will not adversely impact the deployment
size of all but the smallest applications or components.

The relatively large comment/code ratio is due to two reasons.  First, we
use \emph{semantic properties} for program
specification~\cite{Kiniry02-SC}.  Thus, this system witnesses a full
formal specification using both code-external documents, written in the
EBON specification language, as well as in-code, using semantic properties
in Javadoc-style comments.  The other reason for the large numbers of
comments is that we used Jass for contract specification and run-time
testing~\cite{Jass01}.

\subsubsection{Implementation Variations} 

We actually provide two versions of IDebug; one for rigorous testing and
one for shipped code.

The first version is the standard package that is ready for delivery.  It
has all assertions turned off, no contracts enabled, etc.  This is the
version that we reported on above with respect to code size.  This version
is assembled in the \texttt{idebug} package.

The second, called \emph{IDebug High-Confidence}, is a version of the
framework where we have used Jass to generate generate test code for all
contracts.  This means that the resulting source code, collected in the
\texttt{idebughc} package, as well as the compiled class files, are
significantly larger that the non-augmented version.  In fact, it is nearly
twice the size of the \texttt{idebug} package and is significantly slower
because of the high overhead of all of Jass's run-time assertion analysis.

\subsubsection{Implementation Performance} 

Performance characteristics of the IDebug framework are entirely based upon
the speed of the Java run-time's \texttt{Throwable.\-printStackTrace} method
and \texttt{Hash\-table} and \texttt{String\-Buffer} implementations, since
these classes are at the core of the exception and assertion-handling
mechanisms in IDebug.

The \texttt{Throwable.\-print\-Stack\-Trace} method is important to
performance because, each time a message or assertion guard is triggered,
the IDebug runtime has to determine if it should, in fact, execute the
corresponding output code.  It has to determine which class is currently in
scope, which thread is currently running, etc.  All of this information is
gleaned by parsing the results of a \texttt{printStackTrace} method
call\footnote{We realize that the information in such a stack trace is not
  \emph{guaranteed} by the Java VM specification, but it is the best-effort
  data source that we have currently available.}.

Admittedly, a profile analysis of IDebug could reveal performance
weaknesses. In general, any performance tuning would mean replacing
data structures rather than changing core algorithms, since the
algorithms are highly optimized in the current version.

In general, we believe that performance is not a high-priority issue
in debugging complex systems, especially distributed or
object-oriented ones.  We make this claim for two reasons:
\begin{enumerate}
\item The debugging phase of an implementation should be part of an ordered
  and reasoned test suite, and thus the use of the debugging framework
  should also be logical and methodical.  In other words, rarely will it be
  the case that all threads within a complex application will have all
  their debugging options turned on simultaneously.
\item We believe that debugging statements should not be written by hand or
  statically inserted into program code.  Debug code should be ``tuneable''
  at compile time, not just runtime\footnote{And thus the reason for our
    development of JPP.  See Section~\ref{sec:Complementary_Tools} for more
    information.}, and thus debug framework performance should only matter
  for critical debug paths, of which there should be few.
\end{enumerate}

\subsection{Complementary Tools}
\label{sec:Complementary_Tools}

Static debugging statements can clutter source code, increase object code
size, and reduce execution speed.  We have developed a application called
JPP, the Java Parsing Pre-Processor, that helps avoid this problem by
automatically transforming semantic property-based specifications into test
code~\cite{KiniryCheong98}.  In short, JPP performs transformations of
embedded program specification, in the form of \emph{Design by
  Contract}-like~\cite{Meyer92a} predicates in documentation comments, into
IDebug test code at compile time.

We are also looking into extended other similar Open Source tools so that
they can optionally use an IDebug-based interface.  Such tools include
iContract, Jass, and the JML tools~\cite{LeavensEtal00}.  Of course, our
own EBON tool suite will also support such functionality.

\section{Related Work}
\label{sec:Related_Work}

While several feature-full commercial Java development environments are on
the market, none that we have reviewed come with an integrated debugging
framework like IDebug.  Several have single classes that provide some kind
of logging interface, but the configurability and extensibility of IDebug
are absent.

\begin{itemize}
\item Jakarta's \myhref{http://jakarta.apache.org/log4j/} {Log4J} package
  is the most popular similar framework for system logging.  It supports
  notions of categories and logging messages to a small number of output
  channels, but provides no support for concurrent or distributed systems.
\item JDK 1.4 comes with a logging interface, as specified in
  \myhref{http://jcp.org/aboutJava/communityprocess/review/jsr047/} {JSR
    47}.  It too is quite immature when compared to IDebug, with support
  similar to that of Log4J, with handlers used for output interfaces,
  runtime configuration, and functional areas for categories.  It, too, has
  no support for concurrent or distributed systems monitoring.
\item Several articles in industry magazines such as Java Developer's
  Journal and Java Report have discussed Java debugging frameworks.  All
  support assertions and integrate into the Java exception model, but none
  support even the most basic features of IDebug.
\item Microsoft's Visual J++ comes with a
  \texttt{com.\-ms.\-cfc.\-util.\-Debug} class that provides a simple
  interface to message logging and assertions.  The supplied implementation
  is not extensible, does not support concurrency nor any of the other
  advanced features of IDebug, but does support conditional compilation and
  a system to switch messages on and off at runtime.  Debugging output
  simply goes to the console or to a dialog box.
\end{itemize}

\section{Conclusion}
\label{sec:Conclusion}

IDebug is one of the most advanced debugging frameworks available
today for Java.  It is extremely configurable, supports a wide range
of application types, and, because it is an open system, is extensible
by the developer.

This work was originally accomplished in 1997, prior to any other logging
system being available.  We have since used it in several major complex
systems, both in academia and industry.  In a non-concurrent or distributed
context, it is equivalent to the popular Log4J and the new
\texttt{java.util.logging} package in JDK 1.4. But because neither supports
complex concurrent or distributed systems, we hope that their developers,
or developers of non-Java complex systems, will learn from our experience
and incorporate some of the ideas from this system into their future work.

\subsection{Future Work}
\label{sec:Future_Work}

We encourage developers to extend IDebug.  In particular, we are interested
in alternative implementations of the \texttt{Debug\-Output} and
\texttt{Debug\-Constants} interfaces.  Below, we list a series of
possibilities for output interfaces.
\begin{itemize}
\item \texttt{DebugOutputDB} --- used to log messages to a database via
  JDBC.
\item \texttt{DebugOutputEventSource} --- send messages to arbitrary
  listeners within a Java virtual machine, perhaps as part of a
  compositional JavaBeans-based application.
\item \texttt{DebugOutputFrame} --- to send messages to an arbitrary
  (Swing/AWT) frame within a larger GUI.
\item \texttt{DebugOutputLog} --- to persistently log messages for
  off-line debugging.
\item \texttt{DebugOutputMessager} --- send messages via a
  JMS-conformant messaging infrastructure.
\item \texttt{DebugOutputRemoteEventSource} --- to provide messages as
  distributed events, perhaps as part of a Jini~\cite{JiniArchOverview98}
  application.
\item \texttt{DebugOutputSpace} --- store debugging events in a
  JavaSpace~\cite{JavaSpaces98}.
\end{itemize}

\subsubsection{Other Planned Development}

We also plan on extending the logging subsystem to provide the developer
with the ability to customize the format of log messages.  In particular,
we think that the addition of a time-stamp and thread identification would
be particularly useful.  This extension will be added via
\texttt{Debug\-Context}, thus will be configurable at all levels.

Adding support for the use of arbitrary regular expressions to denote
per-class and per-package monitoring is also of interest.  In general, we
have found the current design adequate, but can see the potential
scalability problems for extremely large-scale applications.

We believe that a debugging GUI that supports both design-time and runtime
customization of \texttt{Debug\-Context} would be useful.  This
functionality would likely be accomplished via a new \texttt{BeanInfo}
subsystem.  Such a GUI would integrate nicely with existing graphical IDEs,
could be used for tutorials, and would be beneficial to non-expert
developers.

Finally, we are investigating integrating IDebug with the {\"U}ber\-Net
distributed messaging infrastructure~\cite{Zimmerman98}. Our primary goal
is to support the currying of call stacks across execution contexts.  This
would mean that assertions and exceptions on remote (receiver) machines
would have access to the call stack of the sending thread.  Currying across
other networking layers, especially RMI, is also of interest to us.

\subsection{Thanks}

The author would like to thank the Infospheres Group for help with the
initial problem analysis and early IDebug design.  In particular, the
comments of Mani Chandy, Dan Zimmerman, Wesley Tanaka, and Adam Rifkin
were invaluable.  Also, Nelson Minar used the first version of IDebug
as part of his thesis work; his comments were very helpful.  Matt
Hanna helped review a previous version of this paper.


\bibliographystyle{plain}



\onecolumn

\appendix

\section{Example Code}
\label{app:Example_Code}

\vspace{1in}

\begin{figure}[h]
  \footnotesize
  \begin{verbatim}
        (1)     public abstract class JikiComponent extends HttpServlet
        (2)       implements Servlet
                {
        (3)       public static Debug debug;
        (4)       public static Assert assert;
        (5)       public static DebugOutput debugOutput;
        (6)       public static JikiIDebugConstants debugConstants;
                  ...etc...
                }
                
        (7)     public class DummyComponent extends JikiComponent
        (8)       implements Servlet
                {
                  ...etc...
        (9)       public DummyComponent()
                  {
        (10)        if (debug == null)
                    {
        (11)          debugConstants = new JikiIDebugConstants();
        (12)          debug = new Debug(debugConstants);
        (13)          assert = debug.getAssert();
        (14)          debugOutput = new ConsoleOutput(debug);
        (15)          debug.setOutputInterface(debugOutput);
                      
                      // turn on debugging if appropriate
        (16)          checkDebugging(debug);
                    }
                  }
                  ...etc...
                }
                
        (17)    public class Dispatcher extends JikiComponent
        (18)      implements Servlet
                {
                  ...etc...
                   *
        (19)       * @precondition ((req != null) && (resp != null)) 
        (20)       * Parameters must be non-null.
                   */
                
        (21)      protected void doGet(HttpServletRequest req, 
        (22)                           HttpServletResponse resp) 
                  {
        (23)        assert.assert(((req != null) && (resp != null)),
        (24)                      "Parameters must be non-null");
                
        (25)        debugOutput.println(debugConstants.TRANSACTION,
        (26)                            "Dispatcher GET: " + req);
                    ...etc...
                  }
                  ...etc...
                }
  \end{verbatim}
  \normalsize
  \caption{Example Code Utilizing the IDebug Framework}
\end{figure}


\end{document}